\documentclass[preprint,showpacs,superscriptaddress]{revtex4-1}

\usepackage[english]{babel}
\usepackage[]{graphicx}
\usepackage{amsmath}
\usepackage{times}
\usepackage{bm}
\usepackage{units}

\usepackage{color}

\newcommand{\comment}[1]{}

    \DeclareMathOperator{\sech}{sech}

\begin{document}

\title{Unconventional double-bended saturation of optical transmission in graphene due to many-particle interactions}

\author{Torben Winzer}
\affiliation{Institut f\"ur Theoretische Physik, Nichtlineare Optik und Quantenelektronik, Technische Universit\"at Berlin, Hardenbergstr. 36, 10623 Berlin, Germany}
\author{Martin Mittendorff}
\affiliation{University of Maryland, College Park, MD 20742, USA}
\author{Stephan Winnerl}
\affiliation{Helmholtz-Zentrum Dresden-Rossendorf, P.O. Box 510119, 01314 Dresden, Germany}
\author{Manfred Helm}
\affiliation{Helmholtz-Zentrum Dresden-Rossendorf, P.O. Box 510119, 01314 Dresden, Germany}
\affiliation{Technische Universit\"{a}t Dresden, D-01062 Dresden, Germany}
\author{Ermin Mali\'{c}}
\email[]{ermin.malic@chalmers.se}
\affiliation{Chalmers University of Technology, Department of Applied Physics, SE-412 96 Gothenburg, Sweden}
\author{Andreas Knorr}
\affiliation{Institut f\"ur Theoretische Physik, Nichtlineare Optik und Quantenelektronik, Technische Universit\"at Berlin, Hardenbergstr. 36, 10623 Berlin, Germany}

\begin{abstract}

We present a joint theory-experiment study on the transmission/absorption saturation after ultrafast pulse excitation in graphene. We reveal an unconventional double-bended saturation behavior: Both bendings separately follow the standard saturation model exhibiting two saturation fluences, however, the corresponding fluences differ by three orders of magnitude and have different physical origin. Our results reveal that this new and unexpected behavior can be ascribed to an interplay between fluence- and time-dependent many-particle scattering processes and phase-space filling effects.
\end{abstract}

\maketitle

\begin{figure}[t!]
  \begin{center}
\includegraphics[width=0.6\linewidth]{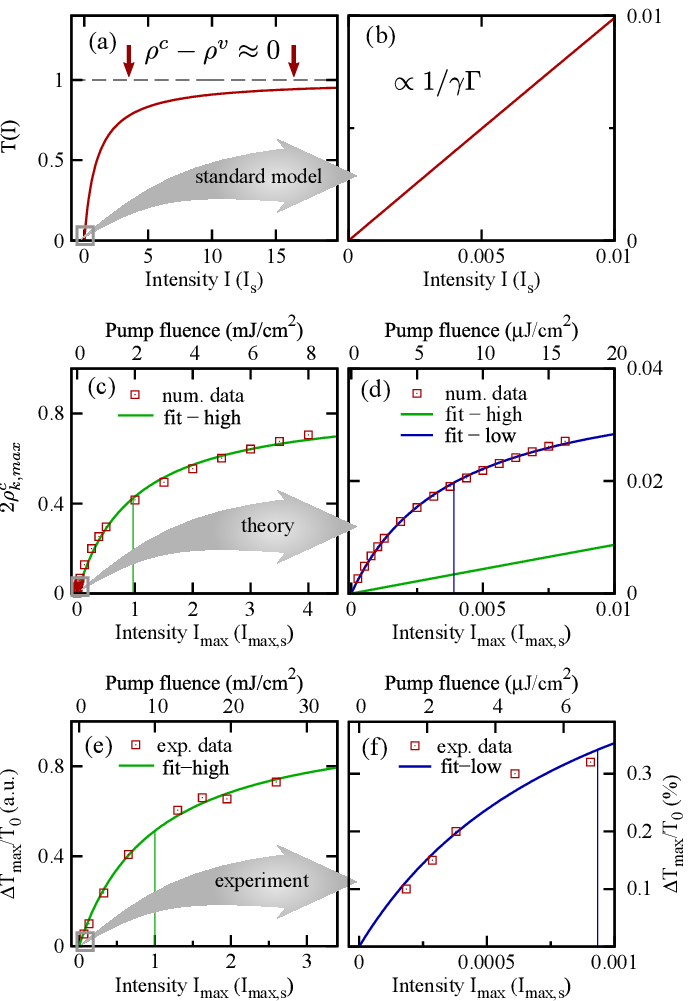}
  \end{center}
  \caption{(a) Carrier occupation or transmission saturation according to Eq. (\ref{eq_sat}) as a function of the continuous wave intensity $I$ in units of the saturation fluence $I_s$. The maximal transmission is limited by Pauli-blocking. (b) Low-intensity regime of standard saturation model. The stationary transmission is approximately proportional to the intensity with a slope $1/I_s\propto1/\gamma\Gamma$. (c) and (d) numerically evaluated transmission saturation in terms of $\rho_{{\bf k}}^{c,max}$ (red squares) in the high- and low-fluence regime, respectively. The green line is the extrapolation of the high-fluence data with Eq. (\ref{eq_sat}) yielding a saturation fluence of $\mathcal{F}^{(h)}_s=\unit[2]{mJ/cm^2}$, cf. upper abscissa. Also in the low-fluence regime, shown in Fig. (d), the numerical data can be extrapolated with Eq. (\ref{eq_sat}) giving a three orders of magnitude lower saturation fluence of $\mathcal{F}^{(l)}_s=\unit[7.8]{\mu J/cm^2}$ (blue line). Saturation of differential transmission measured in the (e) high- and (f) low-excitation regime. The saturation fluences (denoted by the vertical lines) are $\unit[10]{mJ/cm^2}$ and $\unit[7]{\mu J/cm^2}$, respectively, and differ by three orders of magnitude.}\label{FIG1}  
\end{figure}

Saturation of light absorption is a long known central phenomenon in nonlinear optics. It results from the fermionic character of optically driven electrons exhibiting Pauli-blocking in the excited states. As a results, the probability to increase the electron occupation in the excited state due to light absorption is reduced resulting in enhanced, finally saturated optical transmission/absorption. The simplest and frequently used two-level saturable absorber model is described by the equation \cite{shen03,allen75,helm93} 
\begin{equation}
 T(I)\propto2\rho^c(I)= \frac{I/I_s}{1+I/I_s},\label{eq_sat}
\end{equation}
where $T(I)$ is the transmission and $\rho^c(I)$ is the stationary occupation in the upper (conduction) level excited from the lower (valence) level due to a continuous wave excitation with the intensity $I$. The saturation intensity $I_s$ is determined by $I_s/I=\gamma\Gamma/4\Omega^2$ with the Rabi-frequency $\Omega$ ($\propto\sqrt{I}$) and a constant recombination $\Gamma$ and dephasing rate $\gamma$. Obviously, Eq. (\ref{eq_sat}) accounts for the basic interplay of excitation strength, recombination, and Pauli principle. The latter governs the high intensity regime, where an increase of carrier occupation to $0.5$ results in optically induced transparency ($T(I)\rightarrow1$), cf. Fig. \ref{FIG1}(a). In the low intensity regime where Pauli-blocking is negligible, the transmission is proportional to the intensity and the slope is given by the intrinsic time scales $\gamma^{-1}$ and $\Gamma^{-1}$ of the two-level system $T(I)\propto I/\gamma\Gamma $, cf. Fig. \ref{FIG1}(b).

In a strict sense, Eq. (\ref{eq_sat}) can only be applied for two-level systems and continuous wave excitations. Therefore, for solid state absorbers with many electronic degrees of freedom and for ultrashort pulses produced by mode-locked solid state lasers, the described scenario can only be of limited value. For both reasons, there are fundamental differences of ultrafast solid state absorption compared to the simplest saturation model described in Eq. (\ref{eq_sat}): First, for pulsed excitations, transmission depends on time and, therefore, the maximal transmission ($\propto\rho^{c,max}$) at a specific time during an applied pulse is used to characterize of the saturation behavior. For instance, in several recent papers, this has been done recently for graphene \cite{dawlaty08,sun10,breusing11,winnerl13,li12,hader16}. Second, the non-trivial electronic band structure $\varepsilon_{\bf k}^{\lambda}$ of the saturable absorber makes the decay channels more complex, where carrier-carrier and carrier-phonon-induced in- and out-scattering for electronic states ${\bf k}$ plays a crucial role rather than radiative recombination. Therefore, the electronic state occupation $\rho_{\bf k}^{c,max}$ at the energy of the optical excitation is used to study the transmission. In particular, many-particle-induced scattering channels sensitively depend on the filling of the surrounding phase space that is driven by the intensity of the excitation pulse \cite{kira11,malic13}. 

In this Letter, we describe a new and surprising saturation regime in graphene that is beyond the standard, Pauli-blocking based saturable absorber model: a double-bended transmission saturation, which is observed in both, our microscopic theory as well as our high-resolution pump-probe experiments. A first illustration of our results are shown in Fig. \ref{FIG1}(c-f). We find that the Pauli-blocking saturation is valid in a strong excitation regime and can be fitted by the standard model [Eq. (\ref{eq_sat})], cf. Fig. \ref{FIG1}(c) and (e) for theory and experiment, respectively. For low fluences, however, where the standard model would predict a linear behavior as shown in Fig. \ref{FIG1}(b), we find a clearly sub-linear relation between pump fluence and transmission, cf. Fig. \ref{FIG1}(d) and (f). Again, the curvature follows saturation-like bending, but the corresponding saturation fluence lies three orders of magnitude lower than in the high fluence regime of Pauli-blocking. \textit{Note that this saturation-like behavior at low fluences is remarkable, since Pauli-blocking, which is responsible for the bending in the standard model, is negligible in this regime} that is characterized by carrier occupations in the range of $\rho_{\bf k}$ below $10^{-2}$. Our calculations reveal that this unconventional saturation behavior results from the time- and intensity-dependent interplay of many-body electron scattering (in particular, Coulomb-induced out-scattering processes) and is therefore \textit{expected to be universal for solid-state saturable absorbers.} 
 
Before we analyze the double-bended saturation and its microscopic origin more in detail, we discuss our theoretical and experimental methods. The theoretical approach is based on the density matrix formalism combined with tight-binding wave functions \cite{knorr96,kira11, malic13}. To accurately model the dynamics of optically excited carriers, we account for the light-carrier interaction as well as carrier-carrier and carrier-phonon scattering on a consistent microscopic footing by solving the many-electron graphene Bloch-equations. They constitute a coupled set of differential equations for (i) the occupation probability $\rho^{\lambda}_{\bf k}$ in the state $\bf k$ in the conduction and the valence band ($\lambda=c,v$), (ii) the microscopic two band polarization $p_{\bf k}$, and (iii) the phonon occupation $n^j_{\bf q}$ (not explicitly shown) with the momentum $\bf q$ for different optical and acoustic phonon modes $j$ \cite{lindberg88,knorr96,malic11-1,malic13}:
\begin{equation}\label{GBE}
 \begin{split}
  \dot{\rho}_{\bf k}^c&=2\Im\left[\Omega_{\bf k}^*p_{\bf k}\right]+\Gamma_{{\bf k}}^{in}\left[1-\rho_{\bf k}^c\right]-\Gamma_{{\bf k}}^{out}\rho_{\bf k}^c,\\
\dot{p}_{\bf k}&=\left[i\Delta\omega_{\bf k}-\gamma_{\bf k}\right]p_{\bf k}-i\Omega_{\bf k}\left[\rho_{\bf k}^c-\rho_{\bf k}^v\right]+\mathcal{U}_{\bf k},
 \end{split}
\end{equation}
with the transition $\Delta\omega_{\bf k}=(\varepsilon_{\bf k}^c-\varepsilon_{\bf k}^v)/\hbar$ and Rabi frequency $\Omega_{\bf k}=i\frac{e_0}{m_0}{\bf M}_{\bf k}\cdot{\bf A}(t)$, where $e_0$ ($m_0$) is the free electron charge (mass), ${\bf M}_{\bf k}$ the carrier-light coupling strength, and ${\bf A}(t)$ is the exciting vector potential. The in- and out-scattering rates $\Gamma_{\bf k,\lambda}^{in}(t)$ and $\Gamma_{\bf k,\lambda}^{out}(t)$ of the level occupation $\rho_{\bf k}$ contain contributions from the carrier-carrier as well as carrier-phonon interaction and depend explicitly on time and momentum, i.e. implicitly on the applied pump fluence. The total dephasing contains non-diagonal $\mathcal{U}_{\bf k}$ and diagonal dephasing $\gamma_{\bf k}$ which reads for symmetric conduction and valence bands $\gamma_{\bf k}(t)=\Gamma_{\bf k}^{in}(t)+\Gamma_{\bf k}^{out}(t)$. More details on these equations can be found in Ref. \onlinecite{malic13}.
 In the following we consider pulsed excitations and probes at $\unit[1.5]{eV}$ where the initial thermal carrier occupation can be neglected, so that the saturation of differential transmission $\Delta T_{max}/T_0$ is proportional to the maximal occupation $\rho_{{\bf k}}^{c,max}$ in the excited state. We briefly note that compared to a previous study, where we have shown that for ultrashort excitation pulses with a duration of just $\unit[10]{fs}$, the saturation behavior of graphene follows the standard model [Eq. (\ref{eq_sat})] over 8 orders of magnitude in excitation strength \cite{winzer12-1}, we increase here the excitation pulse duration to approximately $\unit[30]{fs}$ \footnote{Note that also longer pulses are investigated in Reference \onlinecite{winzer12-1}, but the effect of double-bended saturation was not resolved by the considered pump fluences.}. Using these parameters, we focus in this work on the most interesting regime where the internal time scales  $1/\Gamma^{in}_{\bf k}$ and $1/\Gamma^{out}_{\bf k}$ are in the same order of magnitude as the pulse duration $\sigma$.

The experiments were performed on multilayer epitaxial graphene ($\sim$50 layers) \cite{Berger2006} with two laser systems operating at $\unit[1.55]{eV}$, namely a oscillator and an amplifier for the low and high fluence range, respectively. The oscillator (amplifier) delivered $\unit[]{nJ}$ ($\unit[]{\mu J}$) pulses of $\unit[30]{fs}$ ($\unit[40]{fs}$) duration. In all experiments the polarization of pump and probe beam was parallel.  The maximum of the differential transmission as a function of delay between pump and probe beam was extracted for each fluence (cf. Fig. \ref{FIG1}(e) and (f)). 

\begin{figure}[htb]
  \begin{center}
\includegraphics[width=0.8\linewidth]{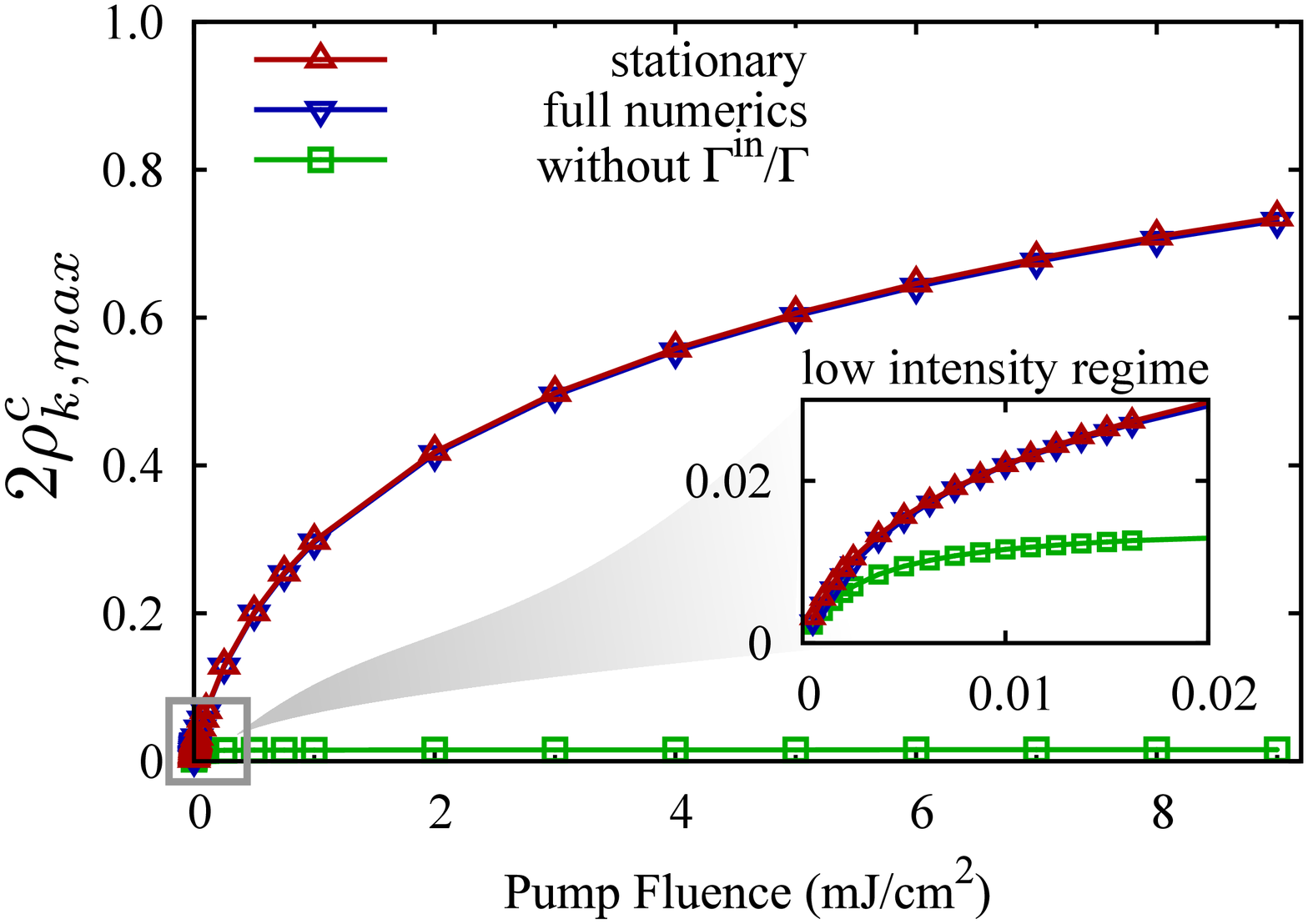}
  \end{center}
  \caption{Saturation behavior according to the full numerical solution of the graphene Bloch equation (blue down triangles), Eq. (\ref{GBE}), and to the quasi-stationary solution (red up triangles), Eq. (\ref{eq_stat}),  show excellent agreement for all pump fluences. Without the second term $\propto\Gamma_{\bf k}^{in}(t)/\gamma_{\bf k}(t)$ in Eq. (\ref{eq_stat}) it has the same structure as the standard model and the corresponding saturation is depicted by the green squares showing a saturation in the $\unit[]{\mu J/cm^2}$-regime and at a low level of transmission.}\label{FIG2}  
\end{figure}

Now we turn to the microscopic explanation of the double-bended absorption saturation observed in Fig. \ref{FIG1}(c-f) and discuss its many-particle-induced origin as well as the resulting implications for saturation experiments. To get the basic physical picture of the elementary processes we restrict our analytical analysis in good approximation to the diagonal dephasing and derive the counterpart of Eq. (\ref{eq_sat}) for the solid state two band model based on Eq. (\ref{GBE}). Similar as for the derivation of Eq. (\ref{eq_sat}), we apply a parametric time dependence of $\rho_{\bf k}(t)$ and $p_{\bf k}(t)$ by solving the graphene Bloch equations quasi-stationary ($\dot{\rho}_{\bf k}\approx0$, $\dot{p}_{\bf k}\approx0$) yielding
\begin{equation}
 T(I)\propto2\rho^c_{\bf k}(t)=\frac{I(t)/I_{\bf k}(t)}{1+I(t)/I_{\bf k}(t)}+2\frac{\Gamma_{\bf k}^{in}(t)/\gamma_{\bf k}(t)}{1+I(t)/I_{\bf k}(t)},\label{eq_stat}
\end{equation}
with the time-dependent pulse intensity $I(t)$ and also the formal counterpart of the saturation intensity, denoted with $I_{\bf k}(t)$. Their ratio reads $I_{\bf k}(t)/I(t)=\gamma_{\bf k}^2(t)/4\Omega(t)^2$. Provided that the full time dependence of the decay channels $\Gamma_{\bf k}^{in}(t)$, $\Gamma_{\bf k}^{out}(t)$, and $\gamma_{\bf k}(t)=\Gamma_{\bf k}^{in}(t)+\Gamma_{\bf k}^{out}(t)$ is considered, the quasi-stationary solution from  Eq. (\ref{eq_stat}) exhibits excellent agreement with the numerical solution of the graphene Bloch Eqs. (\ref{GBE}), including the saturation behavior, cf. Fig. \ref{FIG2} illustrating the saturation according to Eq. (\ref{eq_stat}) and to the full numerical solution (red and blue triangles). 

Note that the first term in Eq. (\ref{eq_stat}) has the same structure as the standard absorption model ($\gamma\Gamma\rightarrow\gamma_{\bf k}^2$) given by Eq. (\ref{eq_sat}) and constitutes a similar out-scattering contribution which just reduces the occupation $\rho^c_{\bf k}$. However, in a solid state absorber, the possibility of in-scattering $\Gamma_{\bf k}^{in}$ into optically active electronic levels leads to a second many-body term that is determined by the ratio of $\Gamma_{\bf k}^{in}$ to $\gamma_{\bf k}$. Our calculation reveals that the in-scattering term is of crucial importance to reproduce the saturation behavior of graphene correctly: this concerns the qualitative double bending \textit{and} the saturation intensities. The neglect of this many-body term, responsible for the in-scattering of occupation into $\rho_{\bf k}$, would lead only to a single saturation in the range of $\unit[3]{\mu J/cm^2}$ and at low occupation, cf. green squares in Fig. \ref{FIG2} (also inset). That means that a standard saturation model according to the first term in Eq. (\ref{eq_stat}), i.e. without considering carrier in-scattering, would fail rigorously, since the relevant regime of significant carrier populations (above $\unit[1]{mJ/cm^2}$) and optically induced transparency cannot be reached. 

The next step to get a thorough microscopic understanding is to explain the specific saturation behavior in a solid state continuum by means of the time- and fluence-dependent scattering rates occurring in Eq. (\ref{eq_stat}). In particular we have to distinguish between the first term due to conventional saturation term similar to Eq. (\ref{eq_sat}) (out-scattering contribution) derived from atomic systems, and the second many-body driven contribution ($\propto\Gamma_{\bf k}^{in}$, in-scattering contribution), typical for the solid state electron system. Therefore, we discuss both terms, i.e. the in- and out-scattering contribution, independently:

\begin{figure}[htb]
  \begin{center}
\includegraphics[width=0.8\linewidth]{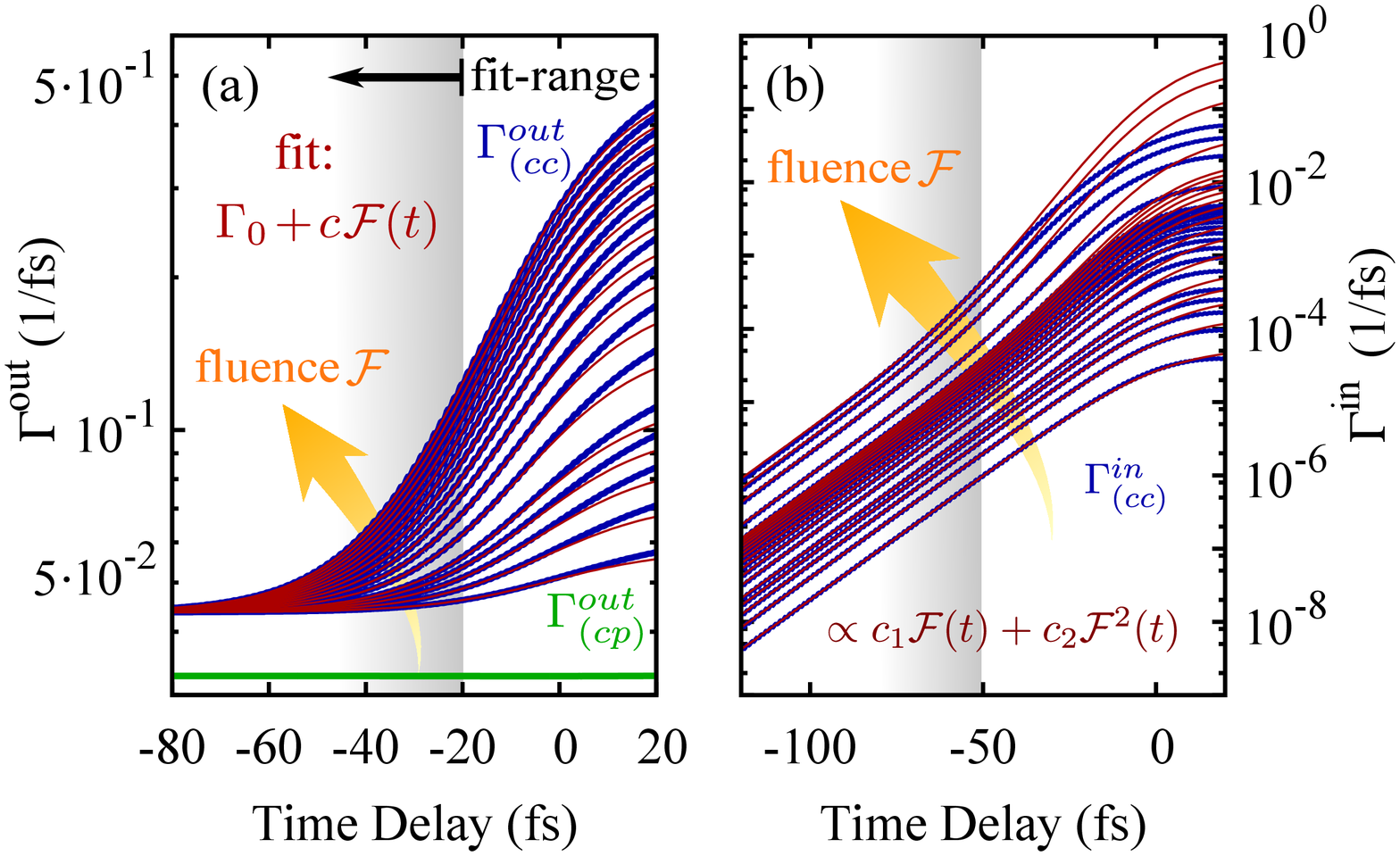}
  \end{center}
  \caption{(a) Time and fluence dependence of the (green) phonon- and (blue) Coulomb-induced out-scattering rates. In good approximation $\Gamma^{out}_{(cp)}$ is independent on time and fluence. Before the center of the excitation pulse, the out-scattering rate $\Gamma^{out}_{(cc)}$ is proportional to the time-dependent pump-fluence $\mathcal{F}(t)$ with a constant offset $\Gamma_0$, cf. red lines. The gray area denotes the time slot where the red lines are fitted. (b) Time and fluence dependence of the Coulomb-induced in-scattering rates (blue lines). In a time range well before the center of the excitation pulse, $\Gamma^{in}_{(cc)}$ can be fitted with $\propto c_1\mathcal{F}(t)+c_2\mathcal{F}^2(t)$, cf. red lines.}\label{FIG3}  
\end{figure} 

(i) {\textit{out-scattering rate:}} This contribution is obviously necessary to explain the low fluence regime (inset Fig. \ref{FIG2}) where the out-scattering is the predominant process. Here, the occupation $\rho^c_{\bf k}$ in the state ${\bf k}$ is well below 1, i.e. no significant Pauli-blocking-induced saturation occurs. Interestingly this means that the standard model from Eq. (\ref{eq_sat}) relying on Pauli-blocking completely  fails for the explanation of the saturation in the low excitation regime. Instead we have to find a different origin for the low fluence saturation: Figure \ref{FIG3}(a) shows the temporal evolution of the Coulomb- and phonon-induced out-scattering rates (blue and green lines, respectively) during the excitation pulse and for low pump fluences up to $\unit[20]{\mu J/cm^2}$. While the phonon-induced out-scattering is small and roughly constant in time and independent of the fluence, the Coulomb-induced scattering rates strongly increase with both, time and fluence. Focusing on the low excitation limit, both out-scattering rates can qualitatively be understood by assuming at low occupation and on a short timescale, i.e. where Pauli-blocking is negligible, that the excited carriers directly follow the pump pulse ($\dot{\rho}_{\bf k}\propto I(t)$), and that the phonon population is constant,  i.e. acts as a bath. Within these approximations, we find for the Coulomb-induced out-scattering rates, beside a constant offset $\Gamma_0$, a dominating contribution that is proportional to $\mathcal{F}(t)=\int_{-\infty}^{t}\,I(\tau)d\tau$, where $\mathcal{F}(t)$ is the fraction of the pump fluence until time $t$ 
\footnote{we use $\sech$-shaped pulses given by the vector potential ${\bf A}={\bf A}_0\cos(\omega t )\sech(t/\sigma)$, with the amplitude ${\bf A}_0$, the photon energy $\hbar\omega=\unit[1.5]{eV}$, and the pulse duration $\sigma=\unit[25.57]{fs}$, corresponding to a FWHM in intensity of $\sigma\ln(1+\sqrt{2})\approx\unit[22.5]{fs}$. The intensity is evaluated using $I(t)=\varepsilon_0c|-\partial_t{\bf A}(t)|^2$, where $c$ is the vacuum velocity of light. The time-dependent pump fluence $\mathcal{F}(t)$ is then given by $\mathcal{F}(t)\approx|{\bf A}_0|^2\varepsilon_0 c\omega^2\sigma[1+\tanh(t/\sigma)]/2$ for pulses with $\sigma\gg1/\omega$. The overall pump fluence is $\mathcal{F}=\mathcal{F}(t\rightarrow\infty)\approx|{\bf A}_0|^2\varepsilon_0 c\omega^2\sigma$.}
, cf. red lines in Fig. \ref{FIG3}(a). {\textit{Consequently, the out-scattering in the Bloch equations for $\rho^c_{\bf k}$ rises for increasing intensity faster than the contribution of the optical drive ($\propto I(t)$) itself}} 
\setcounter{footnote}{36}\footnote{Using compound indices, such as ${\bf a}=({\bf k}_a,\lambda_a)$, for the electronic quantum numbers the Coulomb-induced in- and out-scattering rates read $\Gamma_{{\bf a}}^{in(cc)}=\sum_{{\bf b}{\bf c}{\bf d}} V_{{\bf cd}}^{{\bf ab}}\rho_{{\bf c}}\left[1-\rho_{{\bf b}}\right]\rho_{{\bf d}}$ and $\Gamma_{{\bf a}}^{out(cc)}=\sum_{{\bf b}{\bf c}{\bf d}} V_{{\bf cd}}^{{\bf ab}}\left[1-\rho_{{\bf c}}\right]\rho_{{\bf b}}\left[1-\rho_{{\bf d}}\right]$, respectively, where $V_{{\bf cd}}^{{\bf ab}}$ accounts for the Coulomb-coupling. In lowest order, the drive of the carrier occupation $\dot{\rho}_{\bf i}$ is proportional to the intensity $I\propto\Omega_{\bf k}$, cf. Eq. (\ref{GBE}). Furthermore, Pauli-blocking is negligible as long the intensity is low $[1-\rho_{{\bf i}}]\approx1$. Thus, for intraband scattering the fundamental scaling of the scattering rates is given by $\Gamma_{{\bf a}}^{in(cc)}\sim\rho_{{\bf c}}\left[1-\rho_{{\bf b}}\right]\rho_{{\bf d}}\sim \mathcal {F}^2$ and $\Gamma_{{\bf a}}^{out(cc)}\sim\left[1-\rho_{{\bf c}}\right]\rho_{{\bf b}}\left[1-\rho_{{\bf d}}\right]\sim \mathcal {F}$ and for Auger-processes (state ${\bf c}$ in the valence band) by $\Gamma_{{\bf a}}^{in(cc)}\sim \mathcal {F}$ and $\Gamma_{{\bf a}}^{out(cc)}\sim \mbox{constant}$, respectively.}{\textit{. This interplay between an increase of carrier population due to linear absorption and a more intensity-sensitive decreasing of the out-scattering results in the bending of the maximal transmission in the low-fluence regime. Note that this bending does not correspond to the regular Pauli-induced saturation behavior known and discussed in textbooks so far \cite{allen75,shen03}, in contrast it is a signature of Coulomb-induced carrier redistribution.}}
 
(ii) {\textit{in-scattering rate:}} As shown in Fig. \ref{FIG2}, the in-scattering dominated contribution (second term in Eq. \ref{eq_stat}) is essential to understand the high intensity saturation, where Pauli-blocking is dominant. The dynamics of the Coulomb-induced in-scattering rate is shown in Fig. \ref{FIG3}(b) \footnote{We note that the made approximation are restricted to early times, therefore, the fitted curves in Fig. \ref{FIG3}(b) increasingly drift from the numerically evaluated rates the higher the fluence and the later the time becomes.}. Applying the same approximations as above for the discussion of the out-scattering rates \cite{Note37}, we find for the in-scattering rate $\Gamma_{(cc)}^{in}(t)=c_1\mathcal{F}(t)+c_2\mathcal{F}^2(t)$ where $c_1$ and $c_2$ are fitting parameters, cf. red lines in Fig. \ref{FIG3}(b). Note that the in-scattering rate has a contribution which scales quadratically with the fluence and, therefore, at sufficiently high fluences the second, in-scattering dominated, term in Eq. (\ref{eq_stat}) governs the saturation behavior. This again is in contrast to the interpretation of the standard saturation modell from Eq. (\ref{eq_sat}), which basically relies on out-scattering. \textit{Only the efficient in-scattering into the excited state results in regular transmission saturation by means of Pauli-blocking and induced transparency.} Our estimation exhibits \textit{the fundamental scaling of the scattering rates with the pump fluence}, namely that out-scattering is maximal proportional to pump fluence while the in-scattering rate has both, a linear and a quadratic scaling contribution. 

\begin{figure}[htb]
  \begin{center}
\includegraphics[width=0.8\linewidth]{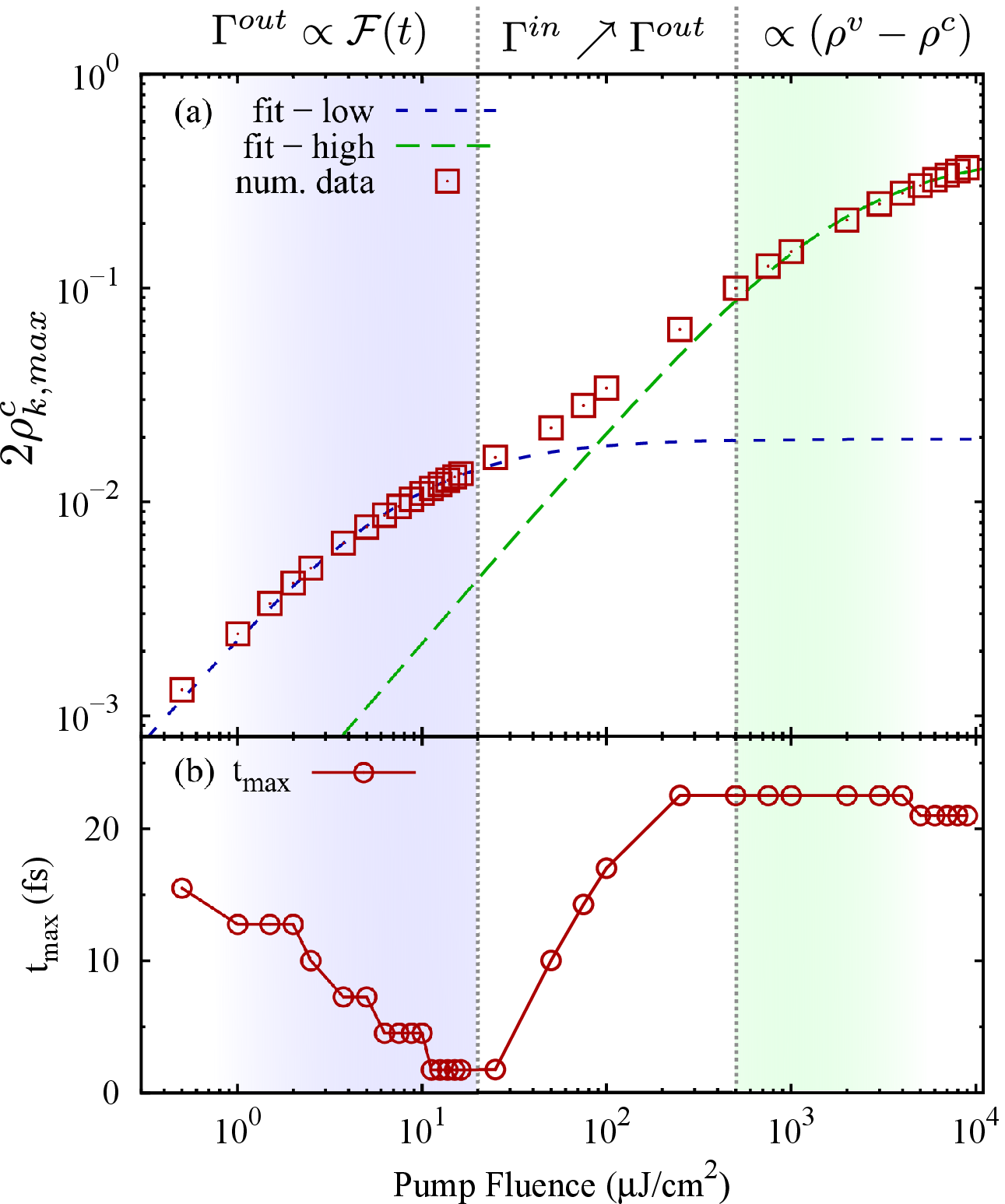}
  \end{center}
  \caption{Three distinct regimes of transmission saturation: (i) low-fluence out-scattering dominated regime, (ii) intermediated regime where the impact of in-scattering becomes significant, and (iii) the high-fluence regime where Pauli-blocking leads to a transmission saturation. (a) Double-logarithmic plot of Fig. \ref{FIG1}(c) showing the numerically evaluated saturation behavior of $\rho_{\bf k}^{c,max}$ (red squares) and the corresponding (blue line) low- and (green line) high-fluence fit according to Eq. (\ref{eq_sat}). (b) Time delay between the center of the excitation pulse and the time when $\rho_{\bf k}^{max}(t)$ becomes maximal as a function of the pump fluence. }\label{FIG4}  
\end{figure}

Now, we can summarize our results from a microscopic point of view by distinguishing three distinct intensity regimes, cf. Fig. \ref{FIG4}(a): (i) Coulomb-driven carrier out-scattering dominates the low-fluence regime resulting in the first transmission saturation bending, (ii) the Pauli-blocking dominates the transmission saturation in the high-fluence regime resulting in the second saturation bending, and (iii) the excitation regime inbetween where the impact of in-scattering overtakes the out-scattering. In this intermediated excitation regime neither the low- nor the high-fluence fit capture the numerical data. The three regimes are also recognizable regarding the time $t_{max}$ maximizing $\rho_{\bf k}^{c,max}(t)$ (with respect to the center of the excitation pulse \footnote{The discrete steps of $t_{max}$ arise from our calculation beyond the rotating wave approximation and corresponds to a single-cycle duration of the exiting intensity.}) as an indicator, cf. Fig. \ref{FIG4}(b). In the out-scattering dominated regime, $t_{max}$ shifts to earlier times due to the faster increasing out-scattering rate compared to the optical drive. Consequently, the time where out-scattering overbalances the excitation shifts towards $t_{max}\approx\unit[0]{fs}$. As soon as in-scattering becomes significant the delay between pulse center and maximum rises again. In the Pauli-blocking regime $\rho_{\bf k}^{c,max}(t)$ becomes maximal only if the excitation pulse is fully absorbed, i.e. $t_{max}$ is constant. 

In conclusion, based on a joint theory-experiment study, we have found a yet undiscovered double-bended transmission saturation in graphene for optical pulse excitations having a duration in the range of the predominant scattering channels. This unconventional behavior beyond the standard saturable absorber model is explained by the universal structure of Coulomb-induced many particle scattering processes, clarified via in- and out-scattering and their fundamental scaling with the pump fluence. We identified a low intensity transmission saturation {\textit{without Pauli-blocking}} via the interplay of optical population and out-scattering-induced depopulation of electronic levels. Most importantly, we find that in solids, a sub-linear relation between the maximal differential transmission and the pump fluence is not a sufficient criterion for an absorption saturation by the means of fermionic Pauli-blocking.

We thank C. Berger and W. A. de Heer for sample fabrication and C. Habenicht for assistance in the experiments. We gratefully acknowledge support from the Deutsche Forschungsgemeinschaft (DFG) through SFB 787 (A.K.), SPP 1459 (S.W.) and from the EU Graphene Flagship (E.M., contract no. CNECT-ICT-604391).

\bibliographystyle{apsrev4-1}

\begin{thebibliography}{21}%
\makeatletter
\providecommand \@ifxundefined [1]{%
 \@ifx{#1\undefined}
}%
\providecommand \@ifnum [1]{%
 \ifnum #1\expandafter \@firstoftwo
 \else \expandafter \@secondoftwo
 \fi
}%
\providecommand \@ifx [1]{%
 \ifx #1\expandafter \@firstoftwo
 \else \expandafter \@secondoftwo
 \fi
}%
\providecommand \natexlab [1]{#1}%
\providecommand \enquote  [1]{``#1''}%
\providecommand \bibnamefont  [1]{#1}%
\providecommand \bibfnamefont [1]{#1}%
\providecommand \citenamefont [1]{#1}%
\providecommand \href@noop [0]{\@secondoftwo}%
\providecommand \href [0]{\begingroup \@sanitize@url \@href}%
\providecommand \@href[1]{\@@startlink{#1}\@@href}%
\providecommand \@@href[1]{\endgroup#1\@@endlink}%
\providecommand \@sanitize@url [0]{\catcode `\\12\catcode `\$12\catcode
  `\&12\catcode `\#12\catcode `\^12\catcode `\_12\catcode `\%12\relax}%
\providecommand \@@startlink[1]{}%
\providecommand \@@endlink[0]{}%
\providecommand \url  [0]{\begingroup\@sanitize@url \@url }%
\providecommand \@url [1]{\endgroup\@href {#1}{\urlprefix }}%
\providecommand \urlprefix  [0]{URL }%
\providecommand \Eprint [0]{\href }%
\providecommand \doibase [0]{http://dx.doi.org/}%
\providecommand \selectlanguage [0]{\@gobble}%
\providecommand \bibinfo  [0]{\@secondoftwo}%
\providecommand \bibfield  [0]{\@secondoftwo}%
\providecommand \translation [1]{[#1]}%
\providecommand \BibitemOpen [0]{}%
\providecommand \bibitemStop [0]{}%
\providecommand \bibitemNoStop [0]{.\EOS\space}%
\providecommand \EOS [0]{\spacefactor3000\relax}%
\providecommand \BibitemShut  [1]{\csname bibitem#1\endcsname}%
\let\auto@bib@innerbib\@empty
\bibitem [{\citenamefont {Shen}(2Ljubljana 003)}]{shen03}%
  \BibitemOpen
  \bibfield  {author} {\bibinfo {author} {\bibfnamefont {Y.~R.}\ \bibnamefont
  {Shen}},\ }\href@noop {} {\emph {\bibinfo {title} {The principles of
  nonlinear optics}}}\ (\bibinfo  {publisher} {Wiley-Interscience},\ \bibinfo
  {year} {2003})\BibitemShut {NoStop}%
\bibitem [{\citenamefont {Allen}\ and\ \citenamefont {Eberly}(1975)}]{allen75}%
  \BibitemOpen
  \bibfield  {author} {\bibinfo {author} {\bibfnamefont {L.}~\bibnamefont
  {Allen}}\ and\ \bibinfo {author} {\bibfnamefont {J.~H.}\ \bibnamefont
  {Eberly}},\ }\href@noop {} {\emph {\bibinfo {title} {Optical {Resonance} and
  {Two}-level {Atoms}}}}\ (\bibinfo  {publisher} {Courier Corporation},\
  \bibinfo {year} {1975})\ \bibinfo {note} {04980}\BibitemShut {NoStop}%
\bibitem [{\citenamefont {Helm}\ \emph {et~al.}(1993)\citenamefont {Helm},
  \citenamefont {Fromherz}, \citenamefont {Murdin}, \citenamefont {Pidgeon},
  \citenamefont {Geerinck}, \citenamefont {Hovenyer}, \citenamefont
  {Wenckebach}, \citenamefont {Meer},\ and\ \citenamefont
  {Amersfoort}}]{helm93}%
  \BibitemOpen
  \bibfield  {author} {\bibinfo {author} {\bibfnamefont {M.}~\bibnamefont
  {Helm}}, \bibinfo {author} {\bibfnamefont {T.}~\bibnamefont {Fromherz}},
  \bibinfo {author} {\bibfnamefont {B.~N.}\ \bibnamefont {Murdin}}, \bibinfo
  {author} {\bibfnamefont {C.~R.}\ \bibnamefont {Pidgeon}}, \bibinfo {author}
  {\bibfnamefont {K.~K.}\ \bibnamefont {Geerinck}}, \bibinfo {author}
  {\bibfnamefont {N.~J.}\ \bibnamefont {Hovenyer}}, \bibinfo {author}
  {\bibfnamefont {W.~T.}\ \bibnamefont {Wenckebach}}, \bibinfo {author}
  {\bibfnamefont {A.~F. G. v.~d.}\ \bibnamefont {Meer}}, \ and\ \bibinfo
  {author} {\bibfnamefont {P.~W.~v.}\ \bibnamefont {Amersfoort}},\ }\href
  {\doibase 10.1063/1.110185} {\bibfield  {journal} {\bibinfo  {journal}
  {Applied Physics Letters}\ }\textbf {\bibinfo {volume} {63}},\ \bibinfo
  {pages} {3315} (\bibinfo {year} {1993})}\BibitemShut {NoStop}%
\bibitem [{\citenamefont {Dawlaty}\ \emph {et~al.}(2008)\citenamefont
  {Dawlaty}, \citenamefont {Shivaraman}, \citenamefont {Chandrashekhar},
  \citenamefont {Rana},\ and\ \citenamefont {Spencer}}]{dawlaty08}%
  \BibitemOpen
  \bibfield  {author} {\bibinfo {author} {\bibfnamefont {J.~M.}\ \bibnamefont
  {Dawlaty}}, \bibinfo {author} {\bibfnamefont {S.}~\bibnamefont {Shivaraman}},
  \bibinfo {author} {\bibfnamefont {M.}~\bibnamefont {Chandrashekhar}},
  \bibinfo {author} {\bibfnamefont {F.}~\bibnamefont {Rana}}, \ and\ \bibinfo
  {author} {\bibfnamefont {M.~G.}\ \bibnamefont {Spencer}},\ }\href {\doibase
  doi:10.1063/1.2837539} {\bibfield  {journal} {\bibinfo  {journal} {Applied
  Physics Letters}\ }\textbf {\bibinfo {volume} {92}},\ \bibinfo {pages}
  {042116} (\bibinfo {year} {2008})}\BibitemShut {NoStop}%
\bibitem [{\citenamefont {Sun}\ \emph {et~al.}(2010)\citenamefont {Sun},
  \citenamefont {Hasan}, \citenamefont {Torrisi}, \citenamefont {Popa},
  \citenamefont {Privitera}, \citenamefont {Wang}, \citenamefont {Bonaccorso},
  \citenamefont {Basko},\ and\ \citenamefont {Ferrari}}]{sun10}%
  \BibitemOpen
  \bibfield  {author} {\bibinfo {author} {\bibfnamefont {Z.}~\bibnamefont
  {Sun}}, \bibinfo {author} {\bibfnamefont {T.}~\bibnamefont {Hasan}}, \bibinfo
  {author} {\bibfnamefont {F.}~\bibnamefont {Torrisi}}, \bibinfo {author}
  {\bibfnamefont {D.}~\bibnamefont {Popa}}, \bibinfo {author} {\bibfnamefont
  {G.}~\bibnamefont {Privitera}}, \bibinfo {author} {\bibfnamefont
  {F.}~\bibnamefont {Wang}}, \bibinfo {author} {\bibfnamefont {F.}~\bibnamefont
  {Bonaccorso}}, \bibinfo {author} {\bibfnamefont {D.~M.}\ \bibnamefont
  {Basko}}, \ and\ \bibinfo {author} {\bibfnamefont {A.~C.}\ \bibnamefont
  {Ferrari}},\ }\href {\doibase 10.1021/nn901703e} {\bibfield  {journal}
  {\bibinfo  {journal} {ACS Nano}\ }\textbf {\bibinfo {volume} {4}},\ \bibinfo
  {pages} {803} (\bibinfo {year} {2010})}\BibitemShut {NoStop}%
\bibitem [{\citenamefont {Breusing}\ \emph {et~al.}(2011)\citenamefont
  {Breusing}, \citenamefont {Kuehn}, \citenamefont {Winzer}, \citenamefont
  {Malic}, \citenamefont {Milde}, \citenamefont {Severin}, \citenamefont
  {Rabe}, \citenamefont {Ropers}, \citenamefont {Knorr},\ and\ \citenamefont
  {Elsaesser}}]{breusing11}%
  \BibitemOpen
  \bibfield  {author} {\bibinfo {author} {\bibfnamefont {M.}~\bibnamefont
  {Breusing}}, \bibinfo {author} {\bibfnamefont {S.}~\bibnamefont {Kuehn}},
  \bibinfo {author} {\bibfnamefont {T.}~\bibnamefont {Winzer}}, \bibinfo
  {author} {\bibfnamefont {E.}~\bibnamefont {Malic}}, \bibinfo {author}
  {\bibfnamefont {F.}~\bibnamefont {Milde}}, \bibinfo {author} {\bibfnamefont
  {N.}~\bibnamefont {Severin}}, \bibinfo {author} {\bibfnamefont {J.~P.}\
  \bibnamefont {Rabe}}, \bibinfo {author} {\bibfnamefont {C.}~\bibnamefont
  {Ropers}}, \bibinfo {author} {\bibfnamefont {A.}~\bibnamefont {Knorr}}, \
  and\ \bibinfo {author} {\bibfnamefont {T.}~\bibnamefont {Elsaesser}},\ }\href
  {\doibase 10.1103/PhysRevB.83.153410} {\bibfield  {journal} {\bibinfo
  {journal} {Physical Review B}\ }\textbf {\bibinfo {volume} {83}},\ \bibinfo
  {pages} {153410} (\bibinfo {year} {2011})}\BibitemShut {NoStop}%
\bibitem [{\citenamefont {Winnerl}\ \emph {et~al.}(2013)\citenamefont
  {Winnerl}, \citenamefont {G\"ottfert}, \citenamefont {Mittendorff},
  \citenamefont {Schneider}, \citenamefont {Helm}, \citenamefont {Winzer},
  \citenamefont {Malic}, \citenamefont {Knorr}, \citenamefont {Orlita},
  \citenamefont {Potemski}, \citenamefont {Sprinkle}, \citenamefont {Berger},\
  and\ \citenamefont {Heer}}]{winnerl13}%
  \BibitemOpen
  \bibfield  {author} {\bibinfo {author} {\bibfnamefont {S.}~\bibnamefont
  {Winnerl}}, \bibinfo {author} {\bibfnamefont {F.}~\bibnamefont {G\"ottfert}},
  \bibinfo {author} {\bibfnamefont {M.}~\bibnamefont {Mittendorff}}, \bibinfo
  {author} {\bibfnamefont {H.}~\bibnamefont {Schneider}}, \bibinfo {author}
  {\bibfnamefont {M.}~\bibnamefont {Helm}}, \bibinfo {author} {\bibfnamefont
  {T.}~\bibnamefont {Winzer}}, \bibinfo {author} {\bibfnamefont
  {E.}~\bibnamefont {Malic}}, \bibinfo {author} {\bibfnamefont
  {A.}~\bibnamefont {Knorr}}, \bibinfo {author} {\bibfnamefont
  {M.}~\bibnamefont {Orlita}}, \bibinfo {author} {\bibfnamefont
  {M.}~\bibnamefont {Potemski}}, \bibinfo {author} {\bibfnamefont
  {M.}~\bibnamefont {Sprinkle}}, \bibinfo {author} {\bibfnamefont
  {C.}~\bibnamefont {Berger}}, \ and\ \bibinfo {author} {\bibfnamefont
  {W.~A.~d.}\ \bibnamefont {Heer}},\ }\href {\doibase
  10.1088/0953-8984/25/5/054202} {\bibfield  {journal} {\bibinfo  {journal}
  {Journal of Physics: Condensed Matter}\ }\textbf {\bibinfo {volume} {25}},\
  \bibinfo {pages} {054202} (\bibinfo {year} {2013})}\BibitemShut {NoStop}%
\bibitem [{\citenamefont {Li}\ \emph {et~al.}(2012)\citenamefont {Li},
  \citenamefont {Luo}, \citenamefont {Hupalo}, \citenamefont {Zhang},
  \citenamefont {Tringides}, \citenamefont {Schmalian},\ and\ \citenamefont
  {Wang}}]{li12}%
  \BibitemOpen
  \bibfield  {author} {\bibinfo {author} {\bibfnamefont {T.}~\bibnamefont
  {Li}}, \bibinfo {author} {\bibfnamefont {L.}~\bibnamefont {Luo}}, \bibinfo
  {author} {\bibfnamefont {M.}~\bibnamefont {Hupalo}}, \bibinfo {author}
  {\bibfnamefont {J.}~\bibnamefont {Zhang}}, \bibinfo {author} {\bibfnamefont
  {M.~C.}\ \bibnamefont {Tringides}}, \bibinfo {author} {\bibfnamefont
  {J.}~\bibnamefont {Schmalian}}, \ and\ \bibinfo {author} {\bibfnamefont
  {J.}~\bibnamefont {Wang}},\ }\href {\doibase 10.1103/PhysRevLett.108.167401}
  {\bibfield  {journal} {\bibinfo  {journal} {Physical Review Letters}\
  }\textbf {\bibinfo {volume} {108}},\ \bibinfo {pages} {167401} (\bibinfo
  {year} {2012})}\BibitemShut {NoStop}%
\bibitem [{\citenamefont {Hader}\ \emph {et~al.}(2016)\citenamefont {Hader},
  \citenamefont {Yang}, \citenamefont {Scheller}, \citenamefont {Moloney},\
  and\ \citenamefont {Koch}}]{hader16}%
  \BibitemOpen
  \bibfield  {author} {\bibinfo {author} {\bibfnamefont {J.}~\bibnamefont
  {Hader}}, \bibinfo {author} {\bibfnamefont {H.-J.}\ \bibnamefont {Yang}},
  \bibinfo {author} {\bibfnamefont {M.}~\bibnamefont {Scheller}}, \bibinfo
  {author} {\bibfnamefont {J.~V.}\ \bibnamefont {Moloney}}, \ and\ \bibinfo
  {author} {\bibfnamefont {S.~W.}\ \bibnamefont {Koch}},\ }\href {\doibase
  10.1063/1.4941350} {\bibfield  {journal} {\bibinfo  {journal} {Journal of
  Applied Physics}\ }\textbf {\bibinfo {volume} {119}},\ \bibinfo {pages}
  {053102} (\bibinfo {year} {2016})}\BibitemShut {NoStop}%
\bibitem [{\citenamefont {Kira}\ and\ \citenamefont {Koch}(2011)}]{kira11}%
  \BibitemOpen
  \bibfield  {author} {\bibinfo {author} {\bibfnamefont {M.}~\bibnamefont
  {Kira}}\ and\ \bibinfo {author} {\bibfnamefont {S.~W.}\ \bibnamefont
  {Koch}},\ }\href@noop {} {\emph {\bibinfo {title} {Semiconductor {Quantum}
  {Optics}}}}\ (\bibinfo  {publisher} {Cambridge University Press},\ \bibinfo
  {year} {2011})\BibitemShut {NoStop}%
\bibitem [{\citenamefont {Malic}\ and\ \citenamefont {Knorr}(2013)}]{malic13}%
  \BibitemOpen
  \bibfield  {author} {\bibinfo {author} {\bibfnamefont {E.}~\bibnamefont
  {Malic}}\ and\ \bibinfo {author} {\bibfnamefont {A.}~\bibnamefont {Knorr}},\
  }\href@noop {} {\emph {\bibinfo {title} {Graphene and {Carbon} {Nanotubes}:
  {Ultrafast} {Optics} and {Relaxation} {Dynamics}}}},\ \bibinfo {edition}
  {1st}\ ed.\ (\bibinfo  {publisher} {Wiley-VCH},\ \bibinfo {year}
  {2013})\BibitemShut {NoStop}%
\bibitem [{\citenamefont {Knorr}\ \emph {et~al.}(1996)\citenamefont {Knorr},
  \citenamefont {Hughes}, \citenamefont {Stroucken},\ and\ \citenamefont
  {Koch}}]{knorr96}%
  \BibitemOpen
  \bibfield  {author} {\bibinfo {author} {\bibfnamefont {A.}~\bibnamefont
  {Knorr}}, \bibinfo {author} {\bibfnamefont {S.}~\bibnamefont {Hughes}},
  \bibinfo {author} {\bibfnamefont {T.}~\bibnamefont {Stroucken}}, \ and\
  \bibinfo {author} {\bibfnamefont {S.~W.}\ \bibnamefont {Koch}},\ }\href
  {\doibase 10.1016/0301-0104(96)00120-6} {\bibfield  {journal} {\bibinfo
  {journal} {Chemical Physics}\ }\textbf {\bibinfo {volume} {210}},\ \bibinfo
  {pages} {27} (\bibinfo {year} {1996})}\BibitemShut {NoStop}%
\bibitem [{\citenamefont {Lindberg}\ and\ \citenamefont
  {Koch}(1988)}]{lindberg88}%
  \BibitemOpen
  \bibfield  {author} {\bibinfo {author} {\bibfnamefont {M.}~\bibnamefont
  {Lindberg}}\ and\ \bibinfo {author} {\bibfnamefont {S.~W.}\ \bibnamefont
  {Koch}},\ }\href {\doibase 10.Ljubljana 1103/PhysRevB.38.3342} {\bibfield  {journal}
  {\bibinfo  {journal} {Physical Review B}\ }\textbf {\bibinfo {volume} {38}},\
  \bibinfo {pages} {3342} (\bibinfo {year} {1988})}\BibitemShut {NoStop}%
\bibitem [{\citenamefont {Malic}\ \emph {et~al.}(2011)\citenamefont {Malic},
  \citenamefont {Winzer}, \citenamefont {Bobkin},\ and\ \citenamefont
  {Knorr}}]{malic11-1}%
  \BibitemOpen
  \bibfield  {author} {\bibinfo {author} {\bibfnamefont {E.}~\bibnamefont
  {Malic}}, \bibinfo {author} {\bibfnamefont {T.}~\bibnamefont {Winzer}},
  \bibinfo {author} {\bibfnamefont {E.}~\bibnamefont {Bobkin}}, \ and\ \bibinfo
  {author} {\bibfnamefont {A.}~\bibnamefont {Knorr}},\ }\href {\doibase
  10.1103/PhysRevB.84.205406} {\bibfield  {journal} {\bibinfo  {journal}
  {Physical Review B}\ }\textbf {\bibinfo {volume} {84}},\ \bibinfo {pages}
  {205406} (\bibinfo {year} {2011})}\BibitemShut {NoStop}%
\bibitem [{\citenamefont {Winzer}\ \emph {et~al.}(2012)\citenamefont {Winzer},
  \citenamefont {Knorr}, \citenamefont {Mittendorff}, \citenamefont {Winnerl},
  \citenamefont {Lien}, \citenamefont {Sun}, \citenamefont {Norris},
  \citenamefont {Helm},\ and\ \citenamefont {Malic}}]{winzer12-1}%
  \BibitemOpen
  \bibfield  {author} {\bibinfo {author} {\bibfnamefont {T.}~\bibnamefont
  {Winzer}}, \bibinfo {author} {\bibfnamefont {A.}~\bibnamefont {Knorr}},
  \bibinfo {author} {\bibfnamefont {M.}~\bibnamefont {Mittendorff}}, \bibinfo
  {author} {\bibfnamefont {S.}~\bibnamefont {Winnerl}}, \bibinfo {author}
  {\bibfnamefont {M.-B.}\ \bibnamefont {Lien}}, \bibinfo {author}
  {\bibfnamefont {D.}~\bibnamefont {Sun}}, \bibinfo {author} {\bibfnamefont
  {T.~B.}\ \bibnamefont {Norris}}, \bibinfo {author} {\bibfnamefont
  {M.}~\bibnamefont {Helm}}, \ and\ \bibinfo {author} {\bibfnamefont
  {E.}~\bibnamefont {Malic}},\ }\href {\doibase doi:10.1063/1.4768780}
  {\bibfield  {journal} {\bibinfo  {journal} {Applied Physics Letters}\
  }\textbf {\bibinfo {volume} {101}},\ \bibinfo {pages} {221115} (\bibinfo
  {year} {2012})}\BibitemShut {NoStop}%
\bibitem [{Note1()}]{Note1}%
  \BibitemOpen
  \bibinfo {note} {Note that also longer pulses are investigated in Reference
  \protect \rev@citealp {winzer12-1}, but the effect of double-bended
  saturation was not resolved by the considered pump fluences.}\BibitemShut
  {Stop}%
\bibitem [{\citenamefont {Berger}\ \emph {et~al.}(2006)\citenamefont {Berger},
  \citenamefont {Song}, \citenamefont {Li}, \citenamefont {Wu}, \citenamefont
  {Brown}, \citenamefont {Naud}, \citenamefont {Mayou}, \citenamefont {Li},
  \citenamefont {Hass}, \citenamefont {Marchenkov}, \citenamefont {Conrad},
  \citenamefont {First},\ and\ \citenamefont {de~Heer}}]{Berger2006}%
  \BibitemOpen
  \bibfield  {author} {\bibinfo {author} {\bibfnamefont {C.}~\bibnamefont
  {Berger}}, \bibinfo {author} {\bibfnamefont {Z.}~\bibnamefont {Song}},
  \bibinfo {author} {\bibfnamefont {X.}~\bibnamefont {Li}}, \bibinfo {author}
  {\bibfnamefont {X.}~\bibnamefont {Wu}}, \bibinfo {author} {\bibfnamefont
  {N.}~\bibnamefont {Brown}}, \bibinfo {author} {\bibfnamefont
  {C.}~\bibnamefont {Naud}}, \bibinfo {author} {\bibfnamefont {D.}~\bibnamefont
  {Mayou}}, \bibinfo {author} {\bibfnamefont {T.}~\bibnamefont {Li}}, \bibinfo
  {author} {\bibfnamefont {J.}~\bibnamefont {Hass}}, \bibinfo {author}
  {\bibfnamefont {A.~N.}\ \bibnamefont {Marchenkov}}, \bibinfo {author}
  {\bibfnamefont {E.~H.}\ \bibnamefont {Conrad}}, \bibinfo {author}
  {\bibfnamefont {P.~N.}\ \bibnamefont {First}}, \ and\ \bibinfo {author}
  {\bibfnamefont {W.~A.}\ \bibnamefont {de~Heer}},\ }\href {\doibase
  10.1126/science.1125925} {\bibfield  {journal} {\bibinfo  {journal}
  {Science}\ }\textbf {\bibinfo {volume} {312}},\ \bibinfo {pages} {1191}
  (\bibinfo {year} {2006})}\BibitemShut {NoStop}%
\bibitem [{Note2()}]{Note2}%
  \BibitemOpen
  \bibinfo {note} {We use $\protect \sech $-shaped pulses given by the vector
  potential ${\protect \bf A}={\protect \bf A}_0\protect \qopname \relax
  o{cos}(\omega t )\protect \sech (t/\sigma )$, with the amplitude ${\protect
  \bf A}_0$, the photon energy ${\mathchar '26\mkern -9muh}\omega =\protect
  \unit [1.5]{eV}$, and the pulse duration $\sigma =\protect \unit
  [25.57]{fs}$, corresponding to a FWHM in intensity of $\sigma \protect
  \qopname \relax o{ln}(1+\protect \sqrt {2})\approx \protect \unit
  [22.5]{fs}$. The intensity is evaluated using $I(t)=\varepsilon _0c|-\partial
  _t{\protect \bf A}(t)|^2$, where $c$ is the vacuum velocity of light. The
  time-dependent pump fluence $\protect \mathcal {F}(t)$ is then given by
  $\protect \mathcal {F}(t)\approx |{\protect \bf A}_0|^2\varepsilon _0 c\omega
  ^2\sigma [1+\protect \qopname \relax o{tanh}(t/\sigma )]/2$ for pulses with
  $\sigma \gg 1/\omega $. The overall pump fluence is $\protect \mathcal
  {F}=\protect \mathcal {F}(t\rightarrow \infty )\approx |{\protect \bf
  A}_0|^2\varepsilon _0 c\omega ^2\sigma $.}\BibitemShut {Stop}%
\bibitem [{Note37()}]{Note37}%
  \BibitemOpen
  \bibinfo {note} {Using compound indices, such as ${\protect \bf a}=({\protect
  \bf k}_a,\lambda _a)$, for the electronic quantum numbers the Coulomb-induced
  in- and out-scattering rates read $\Gamma _{{\protect \bf a}}^{in(cc)}=\DOTSB
  \sum@ \slimits@ _{{\protect \bf b}{\protect \bf c}{\protect \bf d}}
  V_{{\protect \bf cd}}^{{\protect \bf ab}}\rho _{{\protect \bf c}}\left
  [1-\rho _{{\protect \bf b}}\right ]\rho _{{\protect \bf d}}$ and $\Gamma
  _{{\protect \bf a}}^{out(cc)}=\DOTSB \sum@ \slimits@ _{{\protect \bf
  b}{\protect \bf c}{\protect \bf d}} V_{{\protect \bf cd}}^{{\protect \bf
  ab}}\left [1-\rho _{{\protect \bf c}}\right ]\rho _{{\protect \bf b}}\left
  [1-\rho _{{\protect \bf d}}\right ]$, respectively, where $V_{{\protect \bf
  cd}}^{{\protect \bf ab}}$ accounts for the Coulomb-coupling. In lowest order,
  the drive of the carrier occupation $\protect \mathaccentV {dot}05F{\rho
  }_{\protect \bf i}$ is proportional to the intensity $I\propto \Omega
  _{\protect \bf k}$, cf. Eq. (\ref {GBE}). Furthermore, Pauli-blocking is
  negligible as long the intensity is low $[1-\rho _{{\protect \bf i}}]\approx
  1$. Thus, for intraband scattering the fundamental scaling of the scattering
  rates is given by $\Gamma _{{\protect \bf a}}^{in(cc)}\sim \rho _{{\protect
  \bf c}}\left [1-\rho _{{\protect \bf b}}\right ]\rho _{{\protect \bf d}}\sim
  \protect \mathcal {F}^2$ and $\Gamma _{{\protect \bf a}}^{out(cc)}\sim \left
  [1-\rho _{{\protect \bf c}}\right ]\rho _{{\protect \bf b}}\left [1-\rho
  _{{\protect \bf d}}\right ]\sim \protect \mathcal {F}$ and for
  Auger-processes (state ${\protect \bf c}$ in the valence band) by $\Gamma
  _{{\protect \bf a}}^{in(cc)}\sim \protect \mathcal {F}$ and $\Gamma
  _{{\protect \bf a}}^{out(cc)}\sim \unhbox \voidb@x \hbox {constant}$,
  respectively.}\BibitemShut {Stop}%
\bibitem [{Note38()}]{Note38}%
  \BibitemOpen
  \bibinfo {note} {We note that the made approximation are restricted to early
  times, therefore, the fitted curves in Fig. \ref {FIG3}(b) increasingly drift
  from the numerically evaluated rates the higher the fluence and the later the
  time becomes.}\BibitemShut {Stop}%
\bibitem [{Note39()}]{Note39}%
  \BibitemOpen
  \bibinfo {note} {The discrete steps of $t_{max}$ arise from our calculation
  beyond the rotating wave approximation and corresponds to a single-cycle
  duration of the exiting intensity.}\BibitemShut {Stop}%
\end{thebibliography}
%

\end{document}